\documentclass[twocolumn,sigconf,screen]{acmart}

\usepackage[l3]{csvsimple}
\usepackage{cleveref}
\usepackage{caption}
\usepackage{subcaption}
\AtBeginDocument{%
  \providecommand\BibTeX{{%
    \normalfont B\kern-0.5em{\scshape i\kern-0.25em b}\kern-0.8em\TeX}}}

\setcopyright{acmcopyright}
\copyrightyear{2023}
\acmYear{2023}
\acmDOI{XXXXXXX.XXXXXXX}

\acmConference[XXX]{XXX}{2023}{X, XX}

\usepackage{cleveref}
\usepackage{listings}
\usepackage{tikz} 

\newcommand{\topic}{\ensuremath{t}} 
\newcommand{\topicnew}{\ensuremath{\topic_{e}}} 
\newcommand{\topicset}{\ensuremath{\mathcal{T}}} 
\newcommand{\tcluster}{\ensuremath{D}}  
\newcommand{\tclusterfuture}{\tcluster_{future}} 
\newcommand{\tclusterpast}{\tcluster_{past}} 
\newcommand{\trep}{\ensuremath{R}}  
\newcommand{\period}{\ensuremath{p}}  
\newcommand{\timeline}{\ensuremath{P}}  
\newcommand{\embedding}{\ensuremath{emb}}

\newcommand{\pemergence}{\ensuremath{\mathcal{E}}} %

\newcommand{\archive}[0]{\ensuremath{\mathcal{A}}}

\newcommand{\dcites}[0]{\ensuremath{E_\tcluster}}
\newcommand{\tgraph}[0]{\ensuremath{\mathcal{G}}}  

\newcommand{\tnodes}[0]{\ensuremath{\topicset}}
\newcommand{\tcites}[0]{\ensuremath{E_\topicset}}
\newcommand{\tneighbors}[0]{\ensuremath{NeighborTopics}}
\newcommand{\tconnected}[0]{\ensuremath{LinkedTopics}}

\newcommand{\comm}[3]{%
\noindent{{\color{#2}{[[{\bf #1 }#3{\bf\ #1}]]}}}}
\newcommand{\bernd}[1]{\comm{B}{red}{#1}}
\newcommand{\camelia}[1]{\comm{C}{blue}{#1}}
\newcommand{\hub}[1]{\comm{H}{blue}{#1}}
\newcommand{\hamed}[1]{\comm{R}{orange}{#1}}

\renewcommand{\comm}[3]{}

\newcommand{\concept}{}  
\newcommand{\repconcept}{}
\newcommand{\conceptother}{} 
\newcommand{\conceptcomm}{} 
\newcommand{\conceptemerge}{} 
\newcommand{\repconceptemerge}{}

\begin{document}

\renewcommand{\concept}{T767C3}
\renewcommand{\conceptother}{T736C3}
\renewcommand{\conceptcomm}{T1391C3} 
\renewcommand{\conceptemerge}{\conceptother}
\renewcommand{\repconcept}{['ner', 'bioner', 'crf']}
\renewcommand{\repconceptemerge}{['chronic illnesses', 'ehr', 'comorbidities']}

\renewcommand{\concept}{T680C6}
\renewcommand{\conceptother}{T485C6}
\renewcommand{\conceptcomm}{T70C6}
\renewcommand{\conceptemerge}{T661C6}
\renewcommand{\repconcept}{['nearest neighbors', 'knn', 'nearest neighbor']}
\renewcommand{\repconceptemerge}{['glycemic', 'hypoglycemia', 'hyperglycemia']}





\title[ATEM: A Topic Evolution Model for the Detection of Emerging Topics in Scientific Archives]{ATEM: A Topic Evolution Model for the Detection of \\Emerging Topics in Scientific Archives}

\titlenote{\href{http://www-bd.lip6.fr/atem/}{http://www-bd.lip6.fr/atem/}}
\author{Hamed Rahimi}
 \authornote{hamed.rahim@sorbonne-universite.fr} 
 \affiliation{
   \institution{Sorbonne University}
   \city{Paris}
   \country{France}
}

\author{Hubert Naacke}
 \affiliation{
   \institution{Sorbonne University}
   \city{Paris}
   \country{France}
}

\author{Camelia Constantin}
 \affiliation{
   \institution{Sorbonne University}
   \city{Paris}
   \country{France}
}

\author{Bernd Amann}
 \affiliation{
   \institution{Sorbonne University}
   \city{Paris}
   \country{France}
}



\begin{abstract}


This paper presents ATEM, a novel framework for studying topic evolution in scientific archives. ATEM is based on dynamic topic modeling and dynamic graph embedding techniques that explore the dynamics of content and citations of documents within a scientific corpus. ATEM explores a new notion of contextual emergence for the discovery of emerging interdisciplinary research topics based on the dynamics of citation links in topic clusters. Our experiments show that ATEM can efficiently detect emerging cross-disciplinary topics within the DBLP archive of over five million computer science articles.

\end{abstract}



\maketitle
\section{Introduction}
\write18{wget https://cdn.sstatic.net/Img/teams/teams-illo-free-sidebar-promo.svg?v=47faa659a05e}

Our understanding of the world and scientific knowledge is constantly evolving over time. For example, the definition of an atom and its particles has changed over the last centuries. The evolution of science is a continuous process that examines the development of new theories shaped by the collective efforts of scientists through research, experimentation, and analysis. These efforts are driven by a variety of factors, including discoveries, technological advances, and the accumulation of empirical evidence. Understanding the evolution of science has the potential to transform the research landscape by promoting up-to-date knowledge and resource allocation. It also has significant implications for research funding and public policy decisions in academic and industrial settings.

There are various philosophical theories and definitions for science evolution. For example, Popper considers science evolution as a Bayesian inference process that updates the logical possibility of falsification \cite{popper1979objective}, or Kuhn introduces the notion of Darwinian epistemology where science evolution is a Darwinian selection process of theories \cite{kuhn:1962_StructureScientificRevolutions}. These theories try to explain the evolution of scientific domains and describe their relations and interactions with different semantics. Several approaches have been proposed in the literature to analyze science evolution in scientific archives. We categorize these approaches into two classes based on different factors that contribute to the advancement of science: Single-Domain Evolution Analysis and Cross-Domain Evolution Analysis.

Single-Domain (SD) evolution refers to the development of scientific knowledge and methods within a particular discipline independent of external factors, while Cross-Domain (CD) evolution refers to the interaction and cross-fertilization of different scientific disciplines leading to new insights and discoveries. SD evolution attempts to describe the change in conceptual characteristics and attention to scientific topics over time. It is widely studied by dynamic topic models\cite{blei:2006_DynamicTopicModels,abdelrazek:2023_TopicModelingAlgorithms}, which reflect the evolution by discovering latent semantic structures of the documents published in different time periods. Some models also take into account the attention of the scientific community through publication and citation activity to improve the evolution measure and, for example, the identification of emerging topics.  
On the other hand, CD evolution is studied by analyzing the representation and citation-based relationships between dynamic topics. By comparing topic representations in different time periods, it is possible to build structured topic evolution networks and identify evolution patterns like topic merge and split.  Document citation networks contain additional information about other semantic relationships between topics, such as topic influence and information flow. By analyzing the structure of these networks over time, it is possible to study various more complex trends in the interaction between different topics.

\begin{figure*}[htbp]
\includegraphics[width=\textwidth]{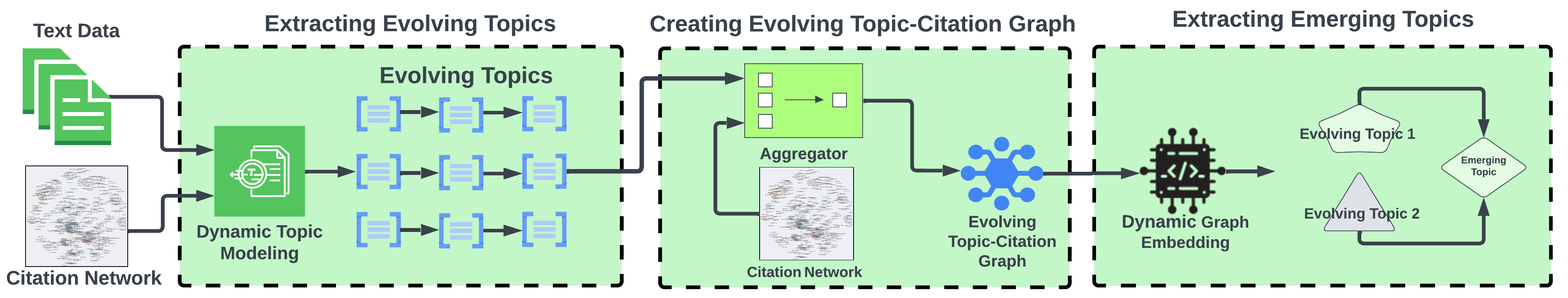}
\caption{The Architecture of ATEM. \normalfont{This framework  extracts evolving topics from a corpus of documents using a dynamic topic model. It then creates a dynamic topic-citation graph by projecting document citation links into the topic space. Finally, it applies a dynamic graph embedding method on the topic-citation graph for representing  the dynamic citation context of topics. Based on this context, emerging topics can be defined as couples or sets of evolving topics with similar citation contexts.}}
\label{fig:arc}
\end{figure*}

One of the most useful analyses of the evolution of science is the detection of topic emergence, which involves the identification of new areas of research and study within scientific disciplines. Emerging topics are ideas or issues that gain attention or become more prominent in a particular field or area of interest. Detecting emerging topics in science has far-reaching implications for society, as it provides a way to track the progression of scientific fields and shape future research and technological development. This task has been described by various communities using different terminologies such as trend analysis~\cite{sharma2016trend}, topic emergence~\cite{small2014identifying}, knowledge flow patterns~\cite{an:2017_IdentifyingDynamicKnowledge}, etc. Existing approaches to detecting emerging scientific themes have their own strengths and limitations. Single-domain-based approaches are generally able to identify trends in the use of specific terms or phrases~\cite{rahimi2023antm}, but may not capture the broader context and relationships between scientific concepts. On the other hand, cross-domain-based approaches can capture the relationships between scientific articles\cite{cordeiro:2018_EvolvingNetworksSocial}, but may be less effective at identifying trends in the usage of specific terms or phrases. 

In this paper, we aim to discover emerging topics by proposing a framework called ATEM that discovers the evolution of science with different analyses. ATEM is motivated by the fact that cross-domain citation links not only indicate semantic relationships between different topics but also indicate a possibility of topic emergence within cited interdisciplinary topics. Dynamic graph embedding allows ATEM to detect emerging topics and predict the new interdisciplinary topics of the future. \Cref{fig:arc} shows the overall architecture of ATEM which will be explained in more detail in \Cref{sec:implementation}.

The rest of the paper is organized as follows. The ATEM science evolution model is introduced in \Cref{sec:atem}. In \Cref{sec:topic-evolution}, we discuss the use of ATEM for various science evolution analysis tasks. The application of ATEM to emergent topic detection is explored in \Cref{sec:detectemerge} and compared to related work in \Cref{sec:relatedwork}. The implementation and the proof of concept are presented in \Cref{sec:implementation} and \Cref{sec:experiments}. Finally, we will conclude in \Cref{sec:conclusion}.

\section{Related Work}
\label{sec:relatedwork}

Topic evolution has  been extensively studied in the context of social media and scholarly  document archives.
Both domains deal with dynamic textual content that is continuously produced by human activity and structured by complex networks connecting people and documents. In the following, we use our definitions of single-domain and cross-domain analysis introduced in the previous section to explore existing work on topic evolution and emergence in scientific archives.

\paragraph{Single-domain evolution analysis}
Single domain topic evolution frameworks track the evolution and emergence of  topics within a given corpus. They exploit different kinds of dynamic information within the document content (e.g., term occurrence and co-occurrence) and the document citation networks (e.g., co-author network, citation links) to detect trends of scientific topics. Topic evolution and trend analysis in science is a very active research area and has been applied in various fields such as business~\cite{rossetto2018structure,zhang2017identify}, information science~\cite{hou:2018_EmergingTrendsNew} or public relations intelligence \cite{santa2018bibliometric}.

Single-domain analysis of \textit{topic content evolution} has been mainly been studied using Dynamic Topic Models. DTM~\cite{blei:2006_DynamicTopicModels} is one of the first  Probabilistic Dynamic Topic Models (PDTM) to incorporate temporal components to study topic evolution. DTM is a variant of LDA~\cite{blei:2003_LatentDirichletAllocation} that creates topics that consist not only of a set of words but also of a timestamp or time range. It uses a Bayesian approach to model the change in the relative proportions of topics within documents over time. DTM is extended by several new models, such as the Discrete-Time Dynamic Topic Model (dDTM)\cite{bahrainian:2017_ModelingDiscreteDynamic} and the Continuous-Time Dynamic Topic Model (cDTM)~\cite{wang:2012_ContinuousTimeDynamic} to handle continuity in different ways. A more recent family of topic models uses novel word embedding and language models to analyze content evolution. For example, Leap2Trend~\cite{dridi:2019_Leap2trendTemporalWorda} relies on temporal word embeddings  to track the dynamics of similarities between pairs of keywords, their rankings, and the respective uprankings (ascents) over time.
%

Hybrid single-domain topic evolution models take advantage of both, the document content  and the structure of citation networks, to analyze topic evolution.
CiteSpace II~\cite{chen:2006_CiteSpaceIIDetecting} is one of the first  systems for detecting and visualizing evolution patterns in scientific archives. It applies various graph analysis methods (burst detection, betweenness centrality) to  analyze the \textit{research front} (transient highly cited publications) within a scientific domain. 
By viewing the evolution of topics as a Markov chain, \cite{zhou:2006_TopicEvolutionSociala} seek to discover the pairwise probabilistic dependency in topics of documents that associate authors from a latent social network. 
\cite{bolelli:2009_FindingTopicTrends} propose a generative model based on LDA for mining distinct topics in document collections by integrating the temporal ordering of documents into the generative process. They integrate the citation graph to boost the weight of  topic terms. 
%
%
%
The Inheritance Topic Model~\cite{he:2009_DetectingTopicEvolutiona} is an iterative topic evolution learning framework that extends LDA with contextual document citation information.
The Emerging Clusters Model~\cite{breitzman:2015_EmergingClustersModel} uses patent citation analysis to asses the impact of patents on subsequent technological developments. The basic idea is to accelerate the detection of emerging topics by analyzing the evolution of citing and cited patents to identify clusters around hot topics. 
In \cite{salatino2017topics}, the authors aim to detect the emergence of new research topics using a hybrid approach based on citation analysis and keywords.
\cite{huang:2021_TrackingDynamicsCoworda} proposes a framework for identifying emerging topics based on dynamic co-word network analysis. Time-sliced co-word networks are weighted according to co-occurrence term frequency, and a back-propagation neural network is used to predict a future co-occurrence term network. 
\cite{morinaga:2004_TrackingDynamicsTopic} propose the usage of a time-stamp based discounting learning algorithm in order to realize real-time topic structure identification for tracking the topic structure by forgetting out-of-date statistics. They apply dynamic model selection for detecting topic emergence by the changes in the finite mixture model.

\bernd{added}
All these recent works demonstrate the efficiency of hybrid methods that combine content-based topic modeling with citation analysis for identifying topic emergence. In our work, we also exploit the citation graph for detecting emerging topics. However, in contrast to hybrid single-domain topic evolution approaches, we apply a cross-domain approach which separates the content-based dynamic topic modeling step from the dynamic citation-network analysis by projecting the document citation graph into the dynamic topic space (see~\Cref{sec:atem}).

\paragraph{Cross-domain evolution analysis}

Cross-domain topic evolution analysis focuses on observing the evolution of topic relationships over time. These relationships can be defined using topic similarity, document co-occurrence within topics, topic citation networks, and also more complex information such as topic influence~\cite{liu:2020_MappingTechnologyEvolutiona} and knowledge flow~\cite{an:2017_IdentifyingDynamicKnowledge}. 
%

A popular way to analyze cross-domain topic evolution is to build topic evolution networks that connect similar dynamic topics from different time periods (similarity is defined by the topic representation). These models make it possible to identify the birth and death of topics, as well as structured evolutionary patterns such as topic splits and merges or paradigm shifts. They can be distinguished first and foremost by the way in which they build up their dynamic topic network.
\cite{jo:2011_WebTopicsDiscoveringa} defines a topic as a unit of evolutionary change in content. Topics are discovered with the time of their appearance in the corpus and connected to form a topic evolution graph using a measure derived from the underlying document network.
\cite{chavalarias:2013_PhylomemeticPatternsScience} defines ''phylomemetic'' networks of topics, where each topic (field) is represented by a cluster of terms that co-occur in a set of documents in a given time period. Fields from different time periods are connected by a matching operator to build a temporal network for identifying emerging, branching, merging, and declining fields. A similar approach was used in~\cite{li:2021_AnalyticGraphData} to propose an extended graph model for querying evolutionary patterns. 
\cite{andrei:2016_ComplexTemporalTopic,beykikhoshk:2018_DiscoveringTopicStructures} use a Hierarchical Dirichlet Process-based (HDP) model and a temporal similarity graph (using Bhattacharyya distance on the topic vector representation) to model complex topic changes.

\cite{jung:2022_IdentifyingCommonPattern,jung:2022_DACDescendantawareClusteringa} also apply a network-based topic evolution approach to represent topics by their past neighbors or ancestors and their structural properties in a temporal sequence of \textit{topic co-occurrence networks}. This representation simultaneously captures content transition and topic correlation and detects complex topic evolution events such as merging, splitting, and emergence. In particular, emerging topics are represented as new nodes, where the context of new topics can be inferred from the contexts and past behavior of their neighboring nodes.
DAC~\cite{jung:2022_DACDescendantawareClusteringa} is a descendant-aware topic clustering algorithm that generates overlapping clusters as candidates for future emerging topics. 
In \cite{jung:2022_IdentifyingCommonPattern}, the authors train binary classification models to capture the materialization of emerging topics.
TermBall~\cite{balili:2020_TermBallTrackingPredicting} builds a weighted dynamic network of co-occurring keywords and extracts topic subgraphs by performing community detection methods. These subgraphs can be used to identify past topic evolution and predict future topic evolution based on structural and temporal features of topic structures.
%
%
The dynamic Bibliographic Knowledge Graph (BKG)~\cite{huo:2022_HotnessPredictionScientific} 
is the result of the merging of bibliographic entities (PMC) with a biomedical and life sciences thesaurus (MeSH). Using different types of topic, paper, author and venue rankings, together with pre-trained node embedding and various pooling techniques, it can be applied to predict topic hotness.
The predictive power of evolutionary topic networks is also validated through the use of community detection algorithms, where the Advanced Clique Percolation Method (ACPM) classification algorithm~\cite{salatino:2018_AUGURForecastingEmergence} has been proposed to identify emerging topic correlations. Topic clusters with notable recent collaborations, found in the semantically enriched topic evolutionary networks and represented by the core publications and author information, are considered as ancestors of a novel topic in its embryonic stage~\cite{salatino:2017_HowAreTopics}.

Our approach to detecting emerging topics also applies the idea of building a dynamic topic citation graph that represents the citation context between topics at different time periods. Using this graph, we apply recent dynamic graph embedding techniques to represent the citation context of topics as embedding vectors for detecting new topics that emerge from sets of topics in a similar context.

\section{ATEM Evolution Model}
\label{sec:atem}

ATEM is a general purpose framework for modeling and analyzing the evolution of topics extracted from scientific archives. ATEM extracts \textit{evolving topics}  using dynamic topic models and build \textit{evolving topic-citation graphs} that connect topics through temporal citation links. 

\subsection{Evolving Topics}
\sloppy
Evolving topics are dynamic topics that describe single-domain evolution of a concept over time with a temporal configuration\cite{rahimi2023antm}. We use the following abstract definition of topics extracted from a scientific archive $\archive$. 
\begin{definition}[Topics]
\label{def:topic}
A topic $\topic\in \topicset$ is a couple $(\tcluster(\topic),\trep(\topic))$ where $\tcluster(\topic)\subseteq \archive$ denotes a subset of \textit{semantically similar} documents, called the document cluster of $\topic$, and $\trep(\topic)$ is a weighted vector of terms in some vocabulary $V$, called the  representation of topic~$\topic$. 
\end{definition}
Observe that this definition can be applied to the topics of any topic model extracted from a document archive $\archive$. Also, the representation $\trep(t)$ of a topic can be computed a priori by the topic model or a posteriori by some labeling function (eg. cTF-IDF\cite{grootendorst:2022_BERTopicNeuralTopica}) on $\tcluster(t)$.

\begin{definition}[Evolving Topics]
\label{def:evolvingtopic}
Given an ordered sequence of possibly overlapping time periods $\timeline=[\period^0, ..., \period^n]$ and a document archive $\archive$, an evolving topic is a sequence of topics $\topic=[\topic^0, ..., \topic^n]$ such that all documents in $\tcluster(\topic^i)$ have been published during the time window $\period^i$ and $\tcluster(t)=\cup_{\topic^i\in \topic} \tcluster(\topic^i)$ is a cluster of \textit{semantically related} documents.
\end{definition}
Observe that there might exist periods $\period^i$ without any sub-topic $\topic^i$ in $\topic$. In this case we suppose that $\tcluster(t^i)$ is empty and $\trep(t^i)$ is undefined. 

Evolving topics can be obtained in two different ways. One possibility is to generate a set of evolving topics $\topicset$ 
by applying a dynamic topic model (e.g. ANTM\cite{rahimi2023antm}, DETM\cite{dieng:2019_DynamicEmbeddedTopic}, BERTopic\cite{grootendorst:2022_BERTopicNeuralTopica}, etc.) on this sequence, and to semantically align the obtained sequence of topic sets $[\topicset^0, ..., \topicset^n]$ into a  set of evolving topics $[\topic^0, ..., \topic^n] \in \topicset^0\times  ...\times \topicset^n$. 
Another possibility is to generate a set of topics over $\archive$ and split each topic cluster $\tcluster(t)$ into a sequence of clusters  $[\tcluster(\topic^0),...,\tcluster(\topic^n)]$ 
where each cluster $\tcluster(\topic^i)\subseteq \tcluster(t)$ is the subset of documents published at period $\period^i$. Remind that the representation $\trep(\topic^i)$ of each sub-topic $\topic^i$ can be obtained a-posteriori from its cluster $\tcluster(\topic^i)$. 
 
\Cref{tab:topicrep-tfidf} shows an example of an evolving topic with the id \concept{} (\Cref{fig:local_evolution_concept}) and the temporal representation of its sub-topics for each year (period) from 2004 to 2018.
\begin{table*}[htbp]
    \caption{Distributed word representation for evolving topic \concept.}
        \label{tab:topicrep-tfidf}
     \centering
\begin{filecontents*}{fig2/T680C6-timeline.csv}
year;concept;label;labeltype;;score
2000;T680C6;['nearest neighbor', 'nearest neighbors', 'classification method', 'neighbors', 'boosting', 'lazy'];tfidf;;
2000;T680C6;[];top2vec;;
2001;T680C6;['nearest neighbor', 'classification tasks', 'nn', 'tangent', 'neighbor', 'committee'];tfidf;;
2001;T680C6;['nearest neighbors', 'nearest neighbour', 'smote', 'knn', 'nearest neighbor', 'statlog'];top2vec;;
2002;T680C6;['mcs', 'nearest neighbor', 'dissimilarity', 'nearest neighbors', 'minimum distance', 'good choice'];tfidf;;
2002;T680C6;['knn', 'nearest neighbors', 'nearest neighbour', 'nearest neighbor', 'classifiers', 'statlog'];top2vec;;
2003;T680C6;['knn', 'nearest neighbors', 'nearest neighbour', 'pop', 'nearest neighbor', 'neighbour'];tfidf;;
2003;T680C6;['knn', 'nearest neighbour', 'nearest neighbors', 'nearest neighbor', 'statlog', 'uci'];top2vec;;
2004;T680C6;['knn', 'linear classifier', 'nearest neighbors', 'nearest neighbor', 'distributional', 'neighbor classifier'];tfidf;;
2004;T680C6;['nearest neighbors', 'knn', 'nearest neighbor', 'nearest neighbour', 'uci', 'classifiers'];top2vec;;
2005;T680C6;['knn', 'nearest neighbors', 'euclidian', 'pearson', 'instance based', 'neighbor nn'];tfidf;;
2005;T680C6;['nearest neighbors', 'knn', 'nearest neighbor', 'nearest neighbour', 'naive bayes', 'smote'];top2vec;;
2006;T680C6;['knn', 'instance based', 'neighbor classifier', 'nearest neighbors', 'neighbor nn', 'neighbor knn'];tfidf;;
2006;T680C6;['knn', 'nearest neighbors', 'nearest neighbor', 'nearest neighbour', 'naive bayes', 'uci'];top2vec;;
2007;T680C6;['knn', 'relief', 'nn classifier', 'neighbor nn', 'membership values', 'nearest neighbors'];tfidf;;
2007;T680C6;['knn', 'nearest neighbors', 'nearest neighbor', 'nearest neighbour', 'naive bayes', 'statlog'];top2vec;;
2008;T680C6;['knn', 'neighbor nn', 'nearest neighbour', 'nn algorithm', 'nearest neighbors', 'instance based'];tfidf;;
2008;T680C6;['nearest neighbors', 'knn', 'nearest neighbor', 'nearest neighbour', 'uci', 'classifiers'];top2vec;;
2009;T680C6;['knn', 'neighbor knn', 'nn classifier', 'nearest neighbors', 'neighbor nn', 'text classification'];tfidf;;
2009;T680C6;['knn', 'nearest neighbors', 'nearest neighbor', 'nearest neighbour', 'naive bayes', 'classifiers'];top2vec;;
2010;T680C6;['nearest neighbors', 'neighbor classification', 'knn', 'metric learning', 'knn classifier', 'neighbor classifier'];tfidf;;
2010;T680C6;['knn', 'nearest neighbors', 'nearest neighbor', 'nearest neighbour', 'uci', 'classifiers'];top2vec;;
2011;T680C6;['instance selection', 'knn', 'neighbor classifier', 'neighbor classification', 'nearest neighbors', 'instance based'];tfidf;;
2011;T680C6;['nearest neighbors', 'knn', 'nearest neighbor', 'nearest neighbour', 'uci', 'smote'];top2vec;;
2012;T680C6;['knn', 'nearest neighbors', 'neighbor knn', 'test sample', 'instance based', 'nn classifier'];tfidf;;
2012;T680C6;['knn', 'nearest neighbors', 'nearest neighbor', 'nearest neighbour', 'uci', 'naive bayes'];top2vec;;
2013;T680C6;['knn', 'nearest neighbors', 'based nearest', 'knn classifier', 'decision boundary', 'neighbor classifier'];tfidf;;
2013;T680C6;['nearest neighbors', 'knn', 'nearest neighbor', 'nearest neighbour', 'naive bayes', 'lmnn'];top2vec;;
2014;T680C6;['knn', 'metric learning', 'nearest neighbors', 'nn classifier', 'knn classifier', 'neighbor nn'];tfidf;;
2014;T680C6;['nearest neighbors', 'nearest neighbor', 'knn', 'nearest neighbour', 'classifiers', 'naive bayes'];top2vec;;
2015;T680C6;['nn classifier', 'knn classifier', 'knn', 'instance based', 'pmc', 'class label'];tfidf;;
2015;T680C6;['nearest neighbor', 'nearest neighbors', 'knn', 'nearest neighbour', 'lmnn', 'classifiers'];top2vec;;
2016;T680C6;['knn', 'instance selection', 'knn algorithm', 'knn classification', 'dpc', 'nn classification'];tfidf;;
2016;T680C6;['knn', 'nearest neighbors', 'nearest neighbor', 'nearest neighbour', 'classifiers', 'naive bayes'];top2vec;;
2017;T680C6;['knn classifier', 'local mean', 'harmonic mean', 'nearest neighbors', 'knn', 'based nearest'];tfidf;;
2017;T680C6;['nearest neighbors', 'nearest neighbor', 'knn', 'nearest neighbour', 'classifiers', 'uci'];top2vec;;
2018;T680C6;['cent', 'knn classifier', 'knn', 'neighbor method', 'instance selection', 'nearest neighbors'];tfidf;;
2018;T680C6;['nearest neighbors', 'knn', 'nearest neighbor', 'nearest neighbour', 'classifiers', 'uci'];top2vec;;
2019;T680C6;['local mean', 'knn algorithm', 'knn', 'neighbor classifier', 'test sample', 'ignorance'];tfidf;;
2019;T680C6;['nearest neighbors', 'nearest neighbor', 'knn', 'nearest neighbour', 'naive bayes', 'classifiers'];top2vec;;
\end{filecontents*}
\csvreader[
    tabular = c|l,
    filter equal={\csvcoliv}{tfidf},
    range=5-19,
    separator=semicolon,
    table head = \hline Year & Label  \\\hline\hline,
    late after line = \\\hline
]{fig2/T680C6-timeline.csv}{}{\csvcoli & \csvcoliii}
   
\end{table*}

\subsection{Evolving Topic-Citation Graph }

ATEM aims to discover citation relationships among the evolving topics as indicator of cross-domain evolution by projecting the structure of citation network into dynamic topic graph. 

\begin{definition}[Evolving Topic Citations]
Let $\topicset$ be a set of evolving topics defined over a document set $\archive$ and $\dcites(\archive)\subseteq \archive\times \archive$ by a set of citation links defined on $\archive$. Then, the topic clusters $\tcluster(\topic_i)$ all topics  $\topic_i \in \topicset$ and the document citation edges $\dcites$ 
define a set of edges $(\topic_x, \topic_y, j)\in \tcites$ from an evolving topic $\topic_x$ to evolving topic $\topic_y$  if there exists at least one citation  from some document in $\tcluster(\topic_x^j)$ to a document in $\tcluster(\topic_y^k)$ where $0\leq k\leq j$ :

\begin{align}
\label{eq:tcites}
\tcites=\{(\topic_x, \topic_y, j) \mid d\in \tcluster(\topic_x^j), d'\in \tcluster(\topic_y^k), 0\leq k\leq j: \dcites(d,d')\}
\end{align}

\end{definition}
We can add to each edge $(t_x,t_y,i)$ in $\tcites$ a weight $w$ which corresponds, for example, to the number of citations which exist between the documents in $\tcluster(t_x^j)$ and $\tcluster(t_y^k)$, $0\leq k\leq i$.

\section{ATEM Evolution Analysis}
\label{sec:topic-evolution}
The proposed evolution model can be used to analyze the evolution of topics within the two defined classes. The first class is called single-domain evolution analysis and observes the change within the representation of words and contents, the size of topic clusters, and the number of incoming and outgoing citation links. The second class is called cross-domain evolution analysis and explores the evolution of topic relationships defined by the topic citation network. 

\paragraph{Single-Domain Analysis}

Using ATEM, one can analyze the change within the word representation of a single topic. As shown in \Cref{tab:topicrep-tfidf}, observing this change allows us to explore the semantic transformation in our understanding of a single topic by examining the words and phrases that are commonly associated with that topic. Besides, one can identify changes in the way people conceptualize a single topic for discussion and research. This kind of information is useful for researchers and companies seeking to stay up-to-date on the way people think about and discuss a particular topic. In addition, ATEM allows us  to explore the publication and citation trends of individual topics by observing the document cluster size and the incoming and outgoing citations over time.


\begin{example}
\Cref{fig:local_evolution_concept,fig:local_evolution_conceptother} show the evolution of citations to topics \concept{} and \conceptother{} (with at least 5 citations). 

\begin{figure}
\includegraphics[width=0.35\textwidth]{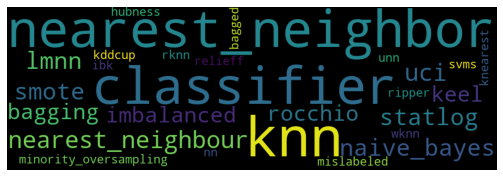}
\includegraphics[width=0.45\textwidth]{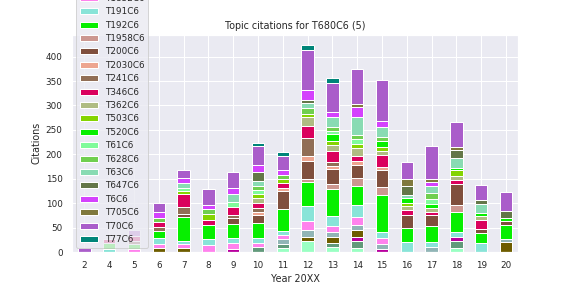}
\caption{Evolution of citations of evolving topic \concept.}
\label{fig:local_evolution_concept}
\end{figure}

\begin{figure}
\includegraphics[width=0.35\textwidth]{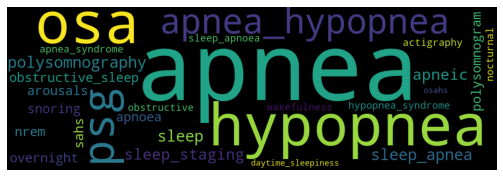}
\includegraphics[width=0.45\textwidth]{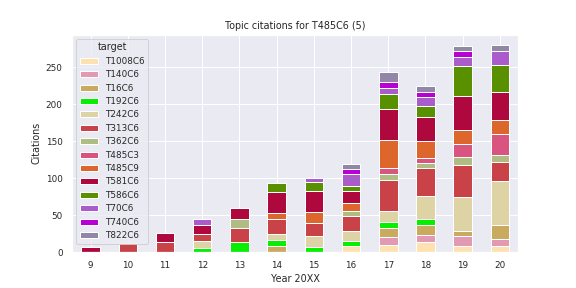}
\caption{Evolution of citations of evolving topic \conceptother}
\label{fig:local_evolution_conceptother}
\end{figure}
\end{example}




\paragraph{Cross-Domain Analysis}

Cross-Domain Evolution Analysis observes the evolution of cross-domain relationships between topics. ATEM investigates factors that contribute to the growth or decline of topics by observing the evolution of citation links between topics. Any variation in the existence and number of citation links between topics is a signal of change for the discovery of new knowledge, and in particular the emergence of new research topics (see \Cref{sec:detectemerge}). 

Citation links between documents can also be used as a way to trace the development and evolution of ideas over time. 
By following the citation links between documents, it is possible to see how different ideas or research findings have been built upon or challenged by subsequent work. 
This can provide insight into the relationships between different topics and how they have evolved over time. 
Therefore, we define a concept called citation context for topics. 
ATEM identifies emerging disciplines by investigating and comparing citation context of different topics.
More exactly, we consider that if two different topics are cited by the same other topics at a given time period, this may indicate the emergence of a new topic defined by the document contents of the cited topics. 
\begin{example}
\Cref{fig:rel_evol1} 
shows the evolution of co-citations to evolving topic \concept{} and \conceptother. We can see that both topics have been cited by topic \conceptcomm{} (\Cref{fig:concomm}) in 2009 and from 2012 to 2020. The number of co-citing topics increases in 2012 which might be considered as a first indication that both concepts are the origin of a new emerging research topic about nearest neighbor classfiers and apnea analysis.

\begin{figure}[htbp]
\includegraphics[width=0.45\textwidth]{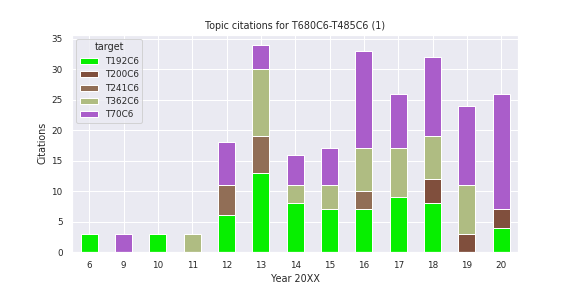}
\caption{Co-citations of \concept{} and \conceptother{}} 
\label{fig:rel_evol1}
\end{figure}

\begin{figure}[htbp]
\includegraphics[width=0.35\textwidth]{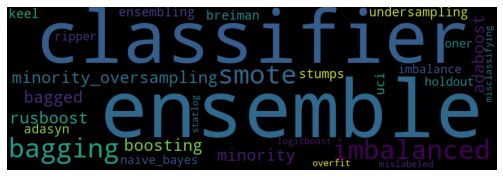}
\caption{Citing evolving topic \conceptcomm{}}
\label{fig:concomm}
\end{figure}
\end{example}

\section{Emerging Topic Detection}
\label{sec:detectemerge}

One of the main goals of ATEM is to identify emerging research topics based on the topic citation graph. Our hypothesis is that citations in documents indicate a relationship between the topics discussed in those documents. When a document cites another document, it typically means that the author of the first document refers to information, ideas, or research from the second document as support or evidence for his or her own work and that the topics discussed in the citing document are influenced or informed by the topics discussed in the second document.  





Citation context refers to the ways in which documents in a given topic cite and are cited by documents in other topics. By analyzing and comparing the citation context of multiple topics, it may then be possible to gain a better understanding of the relationships between different ideas and their evolution over time. In other words, \textit{if two or more topics share a high citation context similarity at a given time period, it is reasonable to expect that they might merge into a new topic in the future, providing a basis for the development of new and interdisciplinary ideas}. Thus, the citation context similarity between two topics can be seen as a measure of the likelihood of discovering new interdisciplinary topics in the future by merging these topics. 



We assume that two evolving topics $\topic_i$ and $\topic_j$ with a \textit{highly similar} 
citation context at a given time period produce an emerging evolving topic $\topic_{i,j}$
Based on this assumption, we need to define a similarity measure that allows us to compare the context of two nodes defined by the topic citation graph $\tcites$. 
A graph embedding~\cite{cai:2018_ComprehensiveSurveyGraph,goyal:2018_GraphEmbeddingTechniquesa} of a graph $G$ is a mapping function $\embedding: G\mapsto 2^{\Re^d}$, which aims to represent nodes, edges, subgraphs, or even the entire graph by low-dimensional feature vectors $v\in \Re^d$ that preserve the topological and other contextual information about the encoded entity. The embedding dimension $d$ is expected to be much smaller than the size of the graph $d\ll n$, where $n$ is the number of nodes in $G$, which allows nodes to be efficiently compared by the encoded properties. 
Using this technique, our idea is then to represent the context of a topic by its embedding in the citation graph and to generate for each evolving topic a \textit{sequence of graph embeddings} which reflects the evolution of its context in the topic citation graph.

There are several dynamic representation learning methods capable of embedding nodes in a low-dimensional vector space which captures the evolution of the network structure. In our implementation, we use dynamic node embeddings~\cite{mahdavi:2019_Dynnode2vecScalableDynamic}, which
project each node $v$ in a sequence of graphs into a \textit{sequence $\embedding(v)$ of low-dimensional vectors}. 
%
By projecting the topic citation links $\tcites$ defined in \cref{sec:atem} on each time period $\period^i\in\timeline$ a set of topic edges $\tcites^i=\{(\topic_x,\topic_y)\mid(\topic_x,\topic_y,i)\in \tcites\}$, we can produce a \textit{sequence of graphs} $\tgraph(\timeline)=[(\tnodes^i,\tcites^i)\mid\period^i\in \timeline]$ ordered by periods $\period^i$ that reflects the distribution of citations between documents of all topics for all time periods. 
Our hypothesis is that the dynamic topic embedding vector $\embedding(\topic)$ of a topic $\topic$ in a dynamic topic citation graph represents the evolution of the citation context of $\topic$, and that two topics with similar embedding (citation context) at period $\period^i$ are likely to generate new emerging topics. More formally, we can now provide a more precise definition of emerging topics:

\begin{definition}[Emerging Topics]
\label{def:emerge2}
Two evolving topics $\topic_i$ and $\topic_j$ define an evolving topic $\topic_{i,j}$ emerging at time period $\period^k$, if the context distance $dist(\topic_i^k,\topic_j^k)$ 
at period $\period^k$ is above a given threshold $\phi$ and below this threshold before $\period^k$. 
\end{definition}
In our implementation, we use $cosine$-similarity on the topic embeddings to compute the distance between two topic contexts.

Using this definition, we can now detect emerging topics in two ways:
\begin{enumerate}
    \item K-nearest neighbors of a given topic $\topic$: we generate for each evolving topic $t$ and period $\period^i$ a set of nearest neighbors with minimal embedding distance higher than a given threshold.
    
    
    \item Cluster the embeddings of each period: we apply a clustering algorithm on the topic embeddings of each period. Each cluster represents an emerging topic defined by a set of similar topics.
    
\end{enumerate}

Note that the definition of an emerging topic does not include the definition of its topic cluster and representation. This is consistent with the observation that the "future" documents of a topic emerging at a given time are not defined. However, as we show in our experiments, we can define functions to estimate the past and future documents using the topic model or a search engine like Google Scholar.

\begin{example}
\Cref{fig:embedding-heatmap} shows the evolution of topics emerging for topic \concept{} in 2013. We can see, for example, topic \conceptemerge{} appears in 2013 as a near embedding neighbor of evolving topic \concept{} (with a maximal distance of $0.2$). 
\begin{figure}[htbp]
\includegraphics[width=0.5\textwidth]{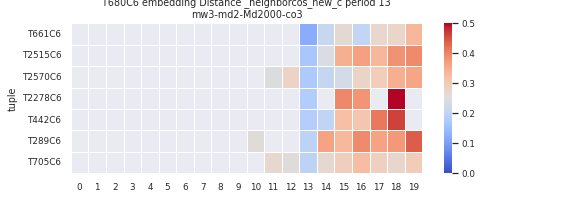}
\caption{Embedding distance evolution for topic \concept{} at period 2013}  
\label{fig:embedding-heatmap}
\end{figure}

\ifthenelse{\equal{\conceptother}{\conceptemerge}}
    {}
    {
\begin{figure}[htbp]
\includegraphics[width=0.35\textwidth]{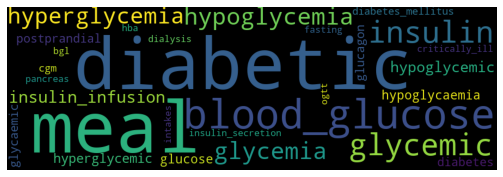}
\caption{Evolving topic \conceptemerge}
\label{fig:wordcloud_conceptemerge}
\end{figure}
}
\Cref{tab:shared_documents} shows the documents that are common to \concept{} and \conceptemerge. These documents are obtained by taking the intersection of 
the results of two queries $\trep(\concept)=$\repconcept{} and $\trep(\conceptemerge)=$\repconceptemerge{} ranked by the average search score. The result shows that most of the top relevant documents for emerging topic (\concept, \conceptemerge) have been published after its emergence period 2013. 
\begin{table*}[htbp]
    \caption{Common documents for topic (\concept,\conceptemerge) emerging in 2013}
        \label{tab:shared_documents}
     \centering
\begin{filecontents*}{fig2/9.csv}
2020.0;Performance evaluation of classification methods with PCA and PSO for diabetes.;0.23403862118721008;0.27683866024017334;3615360
2020.0;An Empirical Evaluation of Machine Learning Techniques for Chronic Kidney Disease Prophecy;0.2699371576309204;0.23124536871910095;3687282
2020.0;Using Machine Learning to Predict the Future Development of Disease;0.22300079464912415;0.2737426459789276;3759429
2019.0;Performance Analysis of Machine Learning Techniques to Predict Diabetes Mellitus.;0.22432051599025726;0.38219690322875977;3453738
2018.0;Accurate Diabetes Risk Stratification Using Machine Learning: Role of Missing Value and Outliers.;0.26091083884239197;0.23068584501743317;3035396
2017.0;Automatic Diagnosis Metabolic Syndrome via a k- Nearest Neighbour Classifier.;0.2626572251319885;0.2674006223678589;2841417
2016.0;Predicting risk of suicide using resting state heart rate.;0.23488864302635193;0.22906461358070374;3325270
2015.0;Computer-aided diagnosis of diabetic subjects by heart rate variability signals using discrete wavelet transform method;0.24455535411834717;0.3306339383125305;1902200
2013.0;Automated detection of diabetes using higher order spectral features extracted from heart rate signals;0.23866119980812073;0.23398324847221375;821224
\end{filecontents*}
\csvreader[
    head = false,
    tabular = c|l,
    separator=semicolon,
    table head = \hline Year & Title  \\\hline\hline,
    late after line = \\\hline
]{fig2/9.csv}{}{\csvcoli & \csvcolii}
   
\end{table*}
\end{example}

\section{Implementation}
\label{sec:implementation}

ATEM is implemented in Python and has been applied to the DBLP\cite{ley:2002_DBLPComputerSciencea} dataset of 5M scientific articles published between 2000 and 2020. DBLP is a large open bibliographic database, search engine, and knowledge graph that archives computer science publications.
%
%
The architecture of ATEM is shown in \Cref{fig:arc}. We can distinguish three main steps that will be explained in the following subsections. 

\subsection{Extracting Evolving Topics}

\begin{figure}[htbp]
\includegraphics[width=0.45\textwidth]{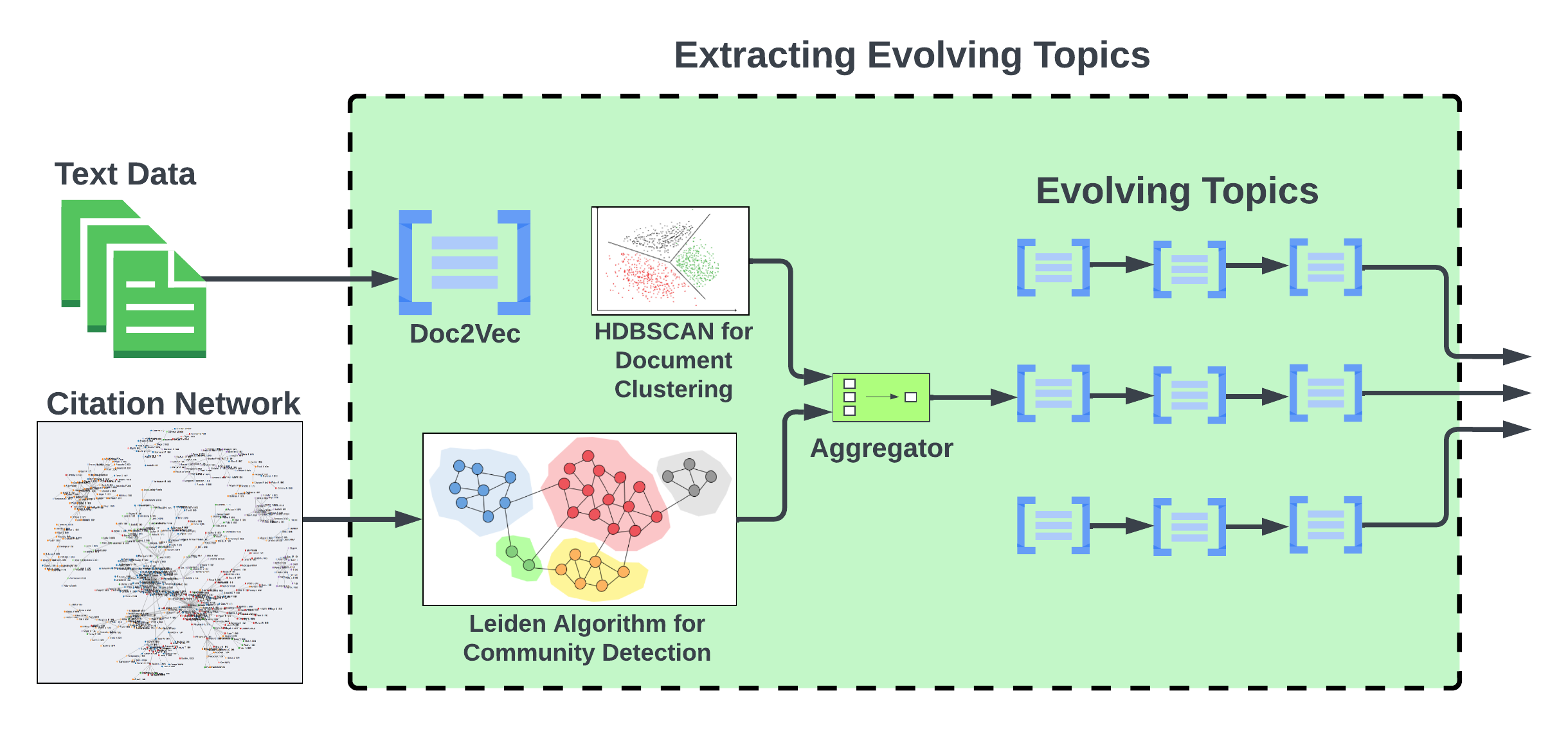}
\caption{The Implementation of Extracting Evolving Topics.}
\label{fig:impevolving}
\end{figure}
The evolving topic extraction workflow consists of four steps, as shown in \Cref{fig:impevolving}.  The first two steps generate two types of document clusters (content-based and citation-based). The first type of cluster corresponds to the traditional notion of topic clusters and regroups semantically similar documents. The second type of cluster applies a community detection algorithm to the document citation graph. We assume that the citation graph reflects different research communities, and by using this process, each generated cluster separates documents published and referenced by different research communities. The third step (cluster aggregation) consists in combining both types of clusters into new clusters corresponding to topic clusters that regroup semantically similar documents published and cited within a scientific community. 
The last step (evolving topic representation) consists in decomposing each cluster into a temporal sequence of sub-clusters (evolving topics) and computing a representation for each sub-cluster.


 
\paragraph{Content-based Document Clustering}
For clustering documents based on their contents, we computed jointly word and document embeddings on the document content  with Doc2Vec~\cite{mikolov:2013_DistributedRepresentationsWords}, then reduced the document dimensions with UMAP~\cite{mcinnes:2018_UmapUniformManifold}, and finally clustered them with HBDSCAN~\cite{campello:2013_DensitybasedClusteringBased}. 
This process is similar to existing cluster-based topic modeling techniques such as Top2Vec~\cite{angelov2020top2vec} and BERTopic~\cite{grootendorst2022bertopic}. 




\paragraph{Citation-based Document Clustering}
For clustering documents based on the citation network, 
we perform the Leiden community detection method~\cite{traag2019louvain} on the citation graph. Each document community then reflects the publication activity within a community of authors that are distinguishable from other author communities in the network. 

\paragraph{Cluster Aggregation} After the two clustering steps, we have two types of document clusters $T$ and $C$ 
which are independently informative and valuable. The cluster aggregation step simply consists in taking the intersection of all clusters in $T$ with all clusters in $C$ to obtain a set of non-empty clusters $\topicset$ where for each cluster $tc_{i,j} \in \topicset$ is the intersection of some cluster $t_i\in T$ and some cluster $c_j\in C$. 


\paragraph{Dynamic Topic Representation} 

In the next step, the topic document clusters $\tcluster\in \topicset$ are divided into $n=|\timeline|$ time frames denoted by $\tcluster=(\tcluster^1,\dots,\tcluster^n)$ where each $\tcluster^i$ is a cluster of documents in period $\period^i$. In this regard, we adopt the dynamic document integration of clusters upon using static time windows. We only keep clusters with a minimal number of $3$ documents.
Each of these topic clusters is represented in two manners:
\begin{itemize}
\item Nearest Words: we compute for each document cluster a centroid vector by averaging over the embeddings of its vectors. The cluster representation is defined by the top-$n$  words corresponding to the $n$ nearest embedding neighbors  of the centroid vector.
\item Class-based TF-IDF: similar to \cite{grootendorst:2022_BERTopicNeuralTopica},   we regroup the documents of each topic cluster $\tcluster(\topic^i)$ in all time periods $\period^i$ and apply TF-IDF to each group to find the top-$n$ word representation for each group. 
\end{itemize}

\subsection{Creating Topic-Citation Graphs}

To generate the evolving topic-citation graph as defined in \Cref{sec:atem}, we create a node for each evolving topic and a weighted directed edge according to \Cref{eq:tcites}. Each citation edge in a given time period is weighted by the number of outgoing citations. \textit{This dynamic representation guarantees that the embedding of a topic in a given period depends only on the citations of the current and previous periods.}

\subsection{Extracting Emerging Topics} 

To compute the temporal node embeddings on the topic citation graph, we used OnlineNode2Vec\cite{beres:2019_NodeEmbeddingsDynamic}, which is based on StreamWalk and online second-order similarity. The result is a temporal embedding for each topic, which can be used to compare evolving topics by their citation context. In particular, it allows us to identify evolving topics $(\topic_x,\topic_y)$ when the distance between two evolving topics $\topic_x$ and $\topic_y$ is less than a given threshold. In our implementation, we use an approximate nearest neighbor algorithm to generate and rank, for each topic $t_x$ and each period, the set of all nearest neighbors $t_y$ that were not nearest neighbors in a previous period. A collection of generated emerging topics is available at \href{https://www.dropbox.com/sh/yxlblk73cx7tv9s/AADtptryp9N83W2PM77DIAiTa?dl=0}{here}.

\section{Proof of concept}
\label{sec:experiments}

\def\version{v24_Md0.2_mn0.2_ms0.22_tkn10_tkd20000_d3_nt100}

The objective of this section is to demonstrate the effectiveness of the proposed framework in identifying emerging topics as compared to co-citation analysis. To achieve this, we categorize the emerging topics into two groups: one based on the embedding representations of the topic-citation graph (referred to as $\tneighbors$), and the other based on co-citation analysis of topics connected through citations (referred to as $\tconnected$). These two groups are then compared to each other. To facilitate this comparison, we generate a set of emerging topics from both $\tneighbors$ and $\tconnected$. We assess the validity of these emerging topics by examining the presence of related documents in the past and future of their discovery. To quantify this, we employ a predictability metric that evaluates the distribution of related documents over time. By scoring the emerging topics based on this metric, we can effectively evaluate their predictive power and performance.

Therefore, we first generate a random sample of 200 evolving topics $T$. For each evolving topic $t\in T$, we generate two sets of $n=10$ topics in each time period $\period^i$:

\begin{enumerate}
    \item $\tneighbors(\topic^i)$ contains $n$ topics $\topic_x$  that are \textit{new} nearest embedding neighbors of $\topic^i$ at period $\period^i$ with a given maximum distance threshold of $0.2$ and minimum embedding norm equal to $0.22$ to remove noisy embedding vectors ($\topic_x$ was not a neighbor before period $\period^i$).
    
    \item $\tconnected(\topic^i)$ contains a random set of $n$ topics $\topic_x$ connected to the evolving topic $\topic$ at period $\period^i$ by a citation path of maximal length equal to $3$.
\end{enumerate}

\sloppy

In each of these sets, each pair $\topicnew=(\topic,\topic_x)$ generated by $\topic_x\in \tconnected(\topic^i)$ and $\topic_x\in \tneighbors(\topic^i)$ is expected to form an emerging topic 
at period $p^i$. 
\bernd{the following is not really exact...} To explore this expectation, we consider all pairs $\topicnew$ as emerging in time period $\period^i$ with representation $\trep(\topicnew)=[\trep(\topic^j)\cup \trep(\topic_x^j) \mid \period^j\in \timeline]$ and a document cluster $\tcluster(\topicnew)=[\tcluster(\topic^j) \cap\allowbreak \tcluster(\topic_x^j)) \mid \period^j\in \timeline]$ over all periods $\period^j$ in $\timeline$.

 We then look at each of these new topics and investigate their emergence predictability based on the year their papers get published. Therefore, we partition $\tcluster(\topicnew)$ into two subsets:  $\tclusterpast(\topicnew)$ of documents published before the emergence period of $\topicnew$, and $\tclusterfuture(\topicnew)$ of documents in $\tcluster(\tcluster)$ published after the emergence period of $\topicnew$. 

 Finally, we quantify the \textit{emergence predictability} $\pemergence$ of each topic pair $\topicnew$ by defining the following function that measures the distribution of its documents before and after its  emergence period:

\begin{equation}
    \pemergence(\topicnew): \frac{|\tclusterfuture(\topicnew)|-|\tclusterpast(\topicnew)|}{|\tcluster(\topicnew)|}
    \label{eq:emergence}
\end{equation}

 meaning (i) when $\pemergence(\topicnew) = 1$, all documents are published at emergence period of $(\topic,\topicnew)$ or afterwards, 
 (ii) when $\pemergence(\topicnew) = 0$, the same number of documents are published before and after the emergence period and 
 (iii) when $\pemergence(\topicnew) = -1$, all documents are published before period $\period$.

While the number of $\tconnected$ increases with time, we observed that the average number of $\tneighbors$ decreases and the average embedding distance increases.

\begin{figure}[htbp]
\includegraphics[width=0.5\textwidth]{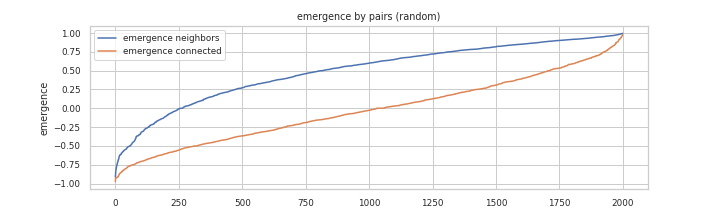}
\caption{The average predictability values of $\tneighbors$ (emergence neighbors) and $\tconnected$ (emergence connected).}
\label{fig:emergence_values_neighbors_and_connected}
\end{figure}

\begin{figure}[htbp]
\includegraphics[width=0.22\textwidth]{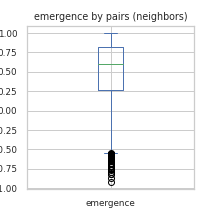}
\includegraphics[width=0.22\textwidth]{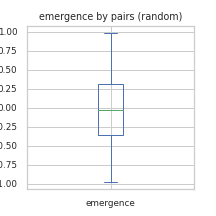}
\caption{Box-plot distribution of predictability values.}
\label{fig:emergence_boxplot}
\end{figure}

\begin{figure}[htbp]
\includegraphics[width=0.22\textwidth]{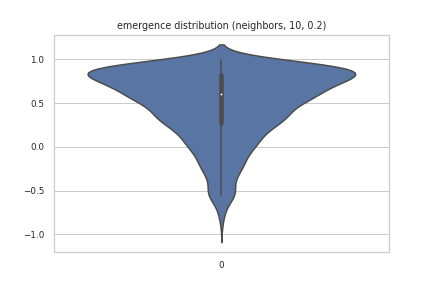}
\includegraphics[width=0.22\textwidth]{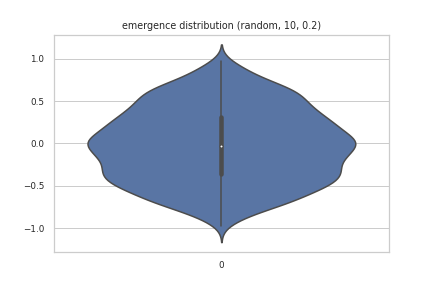}
\caption{Violin distribution of predictability values.}
\label{fig:emergence_violin}
\end{figure}

\Cref{fig:emergence_values_neighbors_and_connected} compares the predictability values for emerging topics of $\tneighbors$ and $\tconnected$. We find that random pairs from $\tneighbors$ have higher predictability compared to $\tconnected$. \Cref{fig:emergence_boxplot,fig:emergence_violin} show the box-plot and violin distribution of predictability values.

%

By \Cref{eq:emergeinv}, we can observe that in average (i) 75\% of emerging topics have $1.25/0.75=1.66$ times more publications after emergence than before and (ii) 50\% of $\tneighbors$, have $2.6/0.4=6.2$ more publications after emergence than before, whereas the ratio is $1$ for $\tconnected$. 
\begin{equation}
\frac{|\tclusterfuture(\topicnew)|}{|\tclusterpast(\topicnew)}|=\frac{\pemergence(\topicnew)+1}{1-\pemergence(\topicnew)}
\label{eq:emergeinv}
\end{equation} 

\begin{figure}[htbp]
\includegraphics[width=0.5\textwidth]{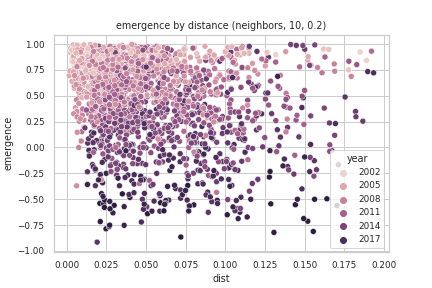}
\caption{The distance distribution of emerging topics.}
\label{fig:emergence_distance}
\end{figure}
\begin{figure}[htbp]
\includegraphics[width=0.5\textwidth]{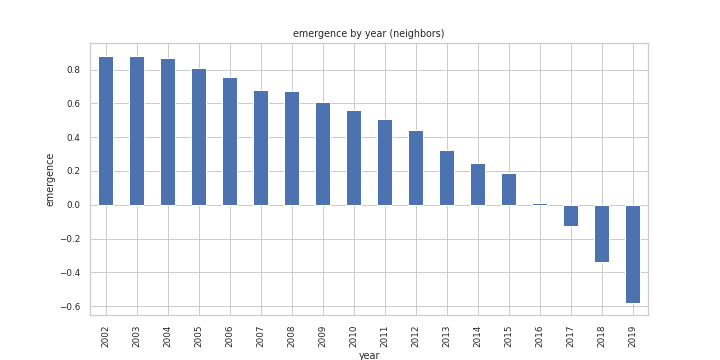}
\caption{The average of predictability values by year}
\label{fig:emergence_by_year}
\end{figure}

\subsection{Emerging topic properties}

\begin{figure}[htbp]
\includegraphics[width=0.4\textwidth]{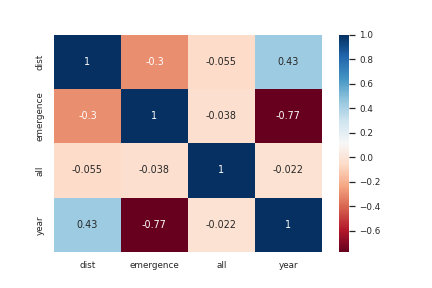}
\caption{The correlation between emerging topic properties}
\label{fig:emergence_distance_correlation}
\end{figure}

\Cref{fig:emergence_distance_correlation} shows the correlation between various parameters that shape the dynamics of emerging topics. We can see that the average embedding distance ($dist$) per period increases in time ($year$). 
\bernd{this is not true; the embeddings of a period do not take account of the future citations!: This can be explained by the fact that as we reach the end of the dataset, we have less contextual information about the future, and correspondingly the average embedding distance increases.} %
This signifies that the applied dynamic embedding method estimates that the analyzed topics get more and more diverse. However this conclusion has to be confirmed by a deeper analysis of the bias introduced by the dynamic computation algorithm.
Second, the predictability ($emergence$) decreases with increasing distance and \textit{strongly decreases} in time (see also \Cref{fig:emergence_distance}).
This is a natural consequence of the definition of emergence which compares the number of relevant documents before and after the emergence period. This number is also influenced by the "relative length" of the past and the future covered by the archive (as shown in \Cref{fig:emergence_by_year}, the average emergence of topic pairs is positive before 2016 and becomes negative afterward). 
Finally, we can see that the average cluster size ($all$) of emerging topics is independent of the period, the average predictability, and the average embedding distance.

\begin{figure}[htbp]
\includegraphics[width=0.4\textwidth]{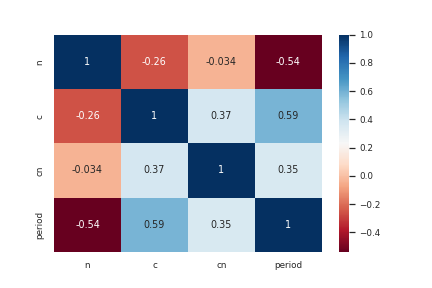}
\caption{The correlations between $\tneighbors$, $\tconnected$ and their intersection.}
\label{fig:emergence_by year}
\end{figure}

\Cref{fig:emergence_by year} shows the correlations between the average number of new emerging topics ($\tneighbors$) by period (n), the average number of randomly connected pairs ($\tconnected$) by period (c), and the average number of connected emerging topics (intersection of $\tneighbors$ and $\tconnected$) by period (cn).
We can see that the average number of embedding neighbors decreases with time, which is consistent with the observation that the embedding distance increases. The fraction of connected neighbors is independent of the number of neighbors, but increases with the number of connected topics. 

\section{Conclusion}
\label{sec:conclusion}
This article presents a new framework for studying the evolution and emergence of topics over time. The analysis framework is based on the notions of single-domain and cross-domain evolution, aiming to distinguish between the evolution of individual topics and the evolution of relationships between topics. This framework is then used to detect emergent topics by using recent graph embedding techniques on topic citation graphs to analyze the evolution of citation context at the topic level and to detect similar topic pairs as new emergent topics. We have implemented this framework, and our experiments show that citation context-based topic similarity is efficient for detecting emerging topics. For future work, we intend to apply our framework to other datasets (e.g., the Microsoft Academic Graph) with different dynamic topic models and to different research domains. 

\section*{Acknowledgement}
We gratefully acknowledge the Sorbonne Center for Artificial Intelligence (SCAI) for partially funding this research through a doctoral fellowship grant. Their support has been instrumental in enabling the successful execution of this study.

\balance 

\bibliographystyle{ieeetr}
\bibliography{biblionew}


\end{document}